\documentclass[12pt,a4paper]{article}
\usepackage[cp1251]{inputenc}
\usepackage[english,russian]{babel}
\usepackage{amssymb}
\begin{document}
\begin{center} {\bfseries Irreducibility of the set of field operators  in Noncommutative Quantum Field
Theory } \vskip 5mm M.~N.~Mnatsakanova$^{\dag}$  and
Yu.~S.~Vernov$^\ddag$ \vskip 5mm {\small {\it $^\dag$ Skobeltsyn
Institute of Nuclear
Physics, Lomonosov Moscow State University, 119992, Vorobyevy Gory, Moscow, Russia}} \\
{\small {\it $^\ddag$ Institute for Nuclear Research of Russian
Academy of Sciences, prospekt 60-letiya Oktyabrya 7a, Moscow
117312, Russia }}
\\
\end{center}
\vskip 5mm \centerline{\bf Abstract}

\begin{abstract}
Irreducibility of the set of quantum field operators has been
proved in noncommutative quantum field theory in the general case
when time does not commute with spatial variables.
\end{abstract}


\section{Introduction}\label{int1}

Irreducibility of the set of quantum field operators
$\varphi\,(x)$ is one of the principal results in axiomatic
quantum field theory (QFT) \cite{SW}, \cite{BLT}.

It implies that, if vacuum vector is cyclic, then the
corresponding set of quantum field operators has to be irreducible
one.

Let us recall that vacuum vector $\Psi_{0}$ is a cyclic one, if
any vector in the space under consideration can be approximated by
a finite linear combination of the vectors
$$
\Psi_{n} = \varphi\,(x_{1}) \ldots  \varphi\,(x_{n})\,\Psi_{0}
$$
with arbitrary accuracy.

In accordance with axiom of  vacuum vector cyclicity any scalar
product in the space in question can be approximated by the linear
combination of Wightman functions
$$
W\,(x_{1}, \dots x_{n}) \equiv \langle\,\Psi_{0}, \varphi\,(x_{1})
\ldots  \varphi\,(x_{n})\,\Psi_{0} \,\rangle.
$$
Let us prove that the set of quantum field operators is
irreducible one in noncommutative quantum field theory (NC QFT) as
well.

Besides, we prove that in usual commutative QFT the irreducibility
of a set of quantum field operators follows from assumptions
weaker then standard.

Let us recall that NC QFT is defined by the Heisenberg-like
commutation relations between coordinates:
% (1)
\begin{equation} \label{cr}
[ \hat{x}_{\mu}, \hat{x}_{\nu}] =i \,\theta_{\mu \nu},
\end{equation}
where $\theta_{\mu\nu}$ is a constant antisymmetric matrix.

It is very important that NC QFT can be also formulated in
commutative space, if we replace the usual product of quantum
field operators (strictly speaking, of the corresponding test
functions) by the $\star$- (Moyal-type) product (see \cite{DN},
\cite{Sz}.

Let us remind that the $\star$-product is defined as
% (2)
$$
\varphi (x) \star\varphi (y) = \exp{\left ({\frac{i}{2} \,
\theta_{\mu\nu} \, \frac{ {\partial}}{\partial x_{\mu}} \, \frac{
{\partial}}{\partial y_{\nu}}} \right)} \,\varphi (x) \varphi (y)
$$
\begin{equation}\label{mprodx}
\equiv \sum^{\infty}_{n = 0}\,\frac{1}{n!}\,{\left ({\frac{i}{2}
\, \theta_{\mu\nu} \, \frac{ {\partial}}{\partial x_{\mu}} \,
\frac{{\partial}}{\partial y_{\nu}}} \right)}^{n} \,\varphi (x)
\varphi (y).
\end{equation}
Evidently the set in equation (\ref{mprodx}) has to be convergent.
It was proved \cite{CMTV7} that this set is a convergent one if
$f\,(x)$ belongs to one of the Gel'fand-Shilov spaces $S^{\beta}$
with $\beta < {1\over 2}$. The similar result was obtained also in
paper \cite{Sol07}.

Noncommutative theories defined by Heisenberg-like commutation
relations (\ref{cr}) can be divided in two classes.

The first of them is the case of only space-space
non-commutativity,  that is $\theta_{0 i} = 0$, time commutes with
spatial coordinates.

It is known that this case is free from the problems with
causality and unitarity  \cite{GM} - \cite{AB} and in this case
the main axiomatic results: CPT and spin-statistics theorems,
Haag's theorem remain valid \cite{AGM} - \cite{CMTV}. Besides,
this case can be obtained as low-energy limit from string theory
\cite{SeWi}.

Let us remind that if time commutes with spatial coordinates, then
there exists one spatial coordinate, say $x_{3}$, which commutes
with all others. Besides it easy to show that this result is valid
in any space if its dimension is even. For simplicity we consider
only four-dimensional case, thus in space-space NC QFT we have two
commuting coordinates and two non-commuting coordinates.

In the second case all coordinates, including time, are
non-commuting.

Let us recall that the set of quantum field operators is
irreducible if from the condition
% (3)
\begin{equation} \label{A}
[A, \, \varphi\,(x)] = 0,
\end{equation}
where $A$ is a bounded operator, follows that
% (4)
\begin{equation} \label{25}
A = C \,\mathbb I \, \qquad C \in \mathbb C  \,
\end{equation}
where $\mathbb I $ is an identity operator.

It is known that if condition (\ref{A}) is fulfilled, then the
space under consideration cannot contain nontrivial subspaces
invariant under the action of the set of operators $\varphi\,(x)$.

Note that actually there is no field operator defined in a point.
Only the smoothed operators written symbolically as
% (5)
\begin{equation} \label{vff}
\varphi_f \equiv\int \,\varphi\,(x) \,f\,(x) \, d \, x,
\end{equation}
where $f\,(x)$ are test functions, can be rigorously defined and
be nontrivial operators in QFT as well as in NC QFT. As the proof
is identical both for operators $\varphi_f$ and $\varphi\,(x)$, so
we give the proof  for $\varphi\,(x)$, for simplicity.

First we consider the case of space-space noncommutativity.

In this case test functions correspond to tempered distributions
in respect with commuting coordinates.

This fact leads to the following spectral condition:
% (6)
\begin{equation} \label{10}
\int \, d \, a \, e^{- i \, p \, a} \, \langle\Phi, U \, (a) \,
\Psi\rangle = 0, \quad \mbox{if} \; p_{0} < \mid p_{3} \mid,
\end{equation}
where $a = \{a_0, a_3 \}$ is a two-dimensional vector, $U \, (a)$
is a translation in the plane $x_0, x_3$, and $\Phi$ and $\Psi$
are arbitrary vectors. The equality (\ref{10}) is similar to the
corresponding equality in the standard case \cite{SW}. It implies
that complete system of physical states (in gauge theories also
nonphysical ones) does not contain tahyon states in momentum space
in respect with commuting coordinates. It means that momentum
$P_n$ for every state satisfies the condition:
$$
P_n^0 \geq | P_n^3 |.
$$

\section{The Proof}\label{sec2}

To prove irreducibility of the set of quantum field operators
fulfilment of the condition
$$
P_n^0 \geq 0
$$
is sufficient.

Let us give the sketch of the proof omitting all technical
details,which are similar to corresponding proof in \cite{SW}.

Let us consider
$$
\langle\, A^{*} \, \Psi_{0}, U \, (a) \, \varphi \,(x_{1}), \ldots
\varphi\,(x_{n}) \,\Psi_{0} \,\rangle.
$$
After simple calculations using the condition (\ref{A}) and
translation operator's unitarity we come to the equality
$$
\langle\, A^{*} \, \Psi_{0}, U \, (a) \, \varphi\,(x_{1}), \ldots
\varphi\,(x_{n}) \,\Psi_{0} \,\rangle =
$$
% (7)
\begin{equation} \label{6}
\langle\, \varphi\,(x_{n})
 \ldots \, \varphi\,(x_{1}), \Psi_{0}, U \, (- a) \,A \, \Psi_{0} \,\rangle.
\end{equation}
In accordance with the spectral condition
$$
\int \, d \, a \, e ^{- i \, p_0 \, a} \, \langle \, A ^*
\,\Psi_0, U\, (a) \, \varphi\,(x_{1})\, \cdots  \,\varphi\,(x_{n})
\,\Psi_0 \,\rangle \neq 0,
$$
only if $p_0 \geq 0$. However,
$$
\int \, d \, a \, e ^{- i \, p_0 \, a} \, \langle \,
\varphi\,_(x_{n})\, \cdots  \,\varphi\,(x_{1}) \,\Psi_0, U \, (-
a) \, A \,\Psi_0 \,\rangle \neq 0,
$$
only if $p_0 \leq 0$. Hence, the equality (\ref{6}) can be
fulfilled only when $p_0 = 0$.

It leads to the equality:
$$
A\,\Psi_0 = C \,\Psi_0, \qquad C\, \in \, \mathbb C
$$
as $\varphi\,(x_{1})\, \cdots  \,\varphi\,(x_{n}) \,\Psi_0 $ is an
arbitrary vector.

From the last equality it follows that
$$
A \, \varphi\,(x_{1})\, \cdots  \,\varphi\,(x_{n}) \Psi_{0} = C \,
\varphi\,(x_{1})\, \cdots  \,\varphi\,(x_{n}) \Psi_{0}.
$$
In order to complete the proof it is necessary to take into
account axiom of vacuum vector cyclicity and the boundedness of
operator $A$.

Now let us show that our statement is true also in the general
case, when time does not commute with spatial variables. The
crucial point in the proof is that
$$
\varphi\,(x)\,\star\,\varphi\,(y) =
\left\{\,\varphi\,(x)\,\star\,\varphi\,(y)\,\right\}_N +
\varepsilon\,(N),
$$
where
$$
\left\{\,\varphi\,(x)\,\star\,\varphi\,(y)\,\right\}_N \equiv
$$
$$
\sum\limits_1^N \,\exp{\left ({\frac{i}{2} \, \theta_{\mu\nu} \,
\frac{ {\partial}}{\partial x_{\mu}} \, \frac{
{\partial}}{\partial y_{\nu}}} \right)} \,\varphi (x) \varphi (y),
$$
and $\varepsilon\,(N) \to 0$ at $N \to \infty$.

At arbitrary $N$ we can derive the statement of the irreducibility
of the set of quantum field operators just as it has been done in
the case of space-space noncommutativity.

As $\varepsilon\,(N) \to 0$ at $N \to \infty$, we can pass to the
limit $N = \infty$.

\section{Conclusions}\label{sec3}

We see that irreducibility of the set of quantum field operators
takes place in the very general theory and under weaker conditions
than usual if axiom of cyclicity of vacuum vector is fulfilled.
The important physical example of such a theory is NC QFT.

\end{document}